\documentstyle[twoside,fleqn,espcrc2]{article}
\newcommand{\dsl}{\not\! \partial}
\newcommand{\eqn}[1]{(\ref{#1})}
\newcommand{\ft}[2]{{\textstyle\frac{#1}{#2}}}
\newcommand  {\Rbar} {{\mbox{\rm$\mbox{I}\!\mbox{R}$}}}
\newcommand {\Cbar}
    {\mathord{\setlength{\unitlength}{1em}
     \begin{picture}(0.6,0.7)(-0.1,0)
        \put(-0.1,0){\rm C}
        \thicklines
        \put(0.2,0.05){\line(0,1){0.55}}
     \end {picture}}}
\newcommand{\cmap}{{\bf c} map}
\newcommand{\rmap}{{\bf r} map}

\newcommand{\Ka}{K\"ahler}
\newcommand{\qu}{quaternionic}

\renewcommand{\d}{\delta}

\newcommand{\del}{\partial}
\def\cF{{\cal F}}

 \def\cM{{\cal M}}
 
\def\cD{{\cal D}}

\def\IZ{{\hbox{{\rm Z}\kern-.4em\hbox{\rm Z}}}}
\def\Im{{\rm Im ~}}
\def\Re{{\rm Re ~}}
\def\bigone{{\hbox{1\kern -.23em{\rm l}}}}
\title{Special geometry and symplectic transformations }
\author{Bernard de Wit\address{Institute for Theoretical Physics,
Utrecht University\\
Princetonplein 5, 3508 TA Utrecht, The Netherlands }%
        and
        Antoine Van Proeyen\address{Instituut voor theoretische fysica\\
 Universiteit Leuven, B-3001 Leuven, Belgium }
        \thanks{Onderzoeksleider, N.F.W.O., Belgium } }
\begin{document}
\setlength{\arraycolsep}{0pt}
\begin{titlepage}
\begin{flushright} THU-95/25\\ KUL-TF-95/32 \\ hep-th/9510186
\end{flushright}
\vfill
\begin{center}
{\large\bf Special geometry and symplectic transformations${}^\dagger$}\\
\vskip 7.mm
{Bernard de Wit }\\
\vskip 0.1cm
{\em Institute for Theoretical Physics} \\
{\em Utrecht University}\\
{\em Princetonplein 5, 3508 TA Utrecht, The Netherlands} \\[5mm]
 Antoine Van Proeyen${}^*$ \\
\vskip 1mm
{\em Instituut voor theoretische fysica}\\
{\em Universiteit Leuven, B-3001 Leuven, Belgium}
\end{center}
\vfill
\begin{center}
{\bf ABSTRACT}
\end{center}
\begin{quote}
Special \Ka\ manifolds are defined by coupling of vector multiplets
to $N=2$ supergravity. The coupling in rigid supersymmetry
exhibits similar features. These models contain
$n$ vectors in rigid supersymmetry and $n+1$ in supergravity, and $n$
complex scalars. Apart from exceptional cases they are defined by
a holomorphic function of the scalars. For supergravity this function is
homogeneous of second degree in an $(n+1)$-dimensional projective
space. Another formulation exists which does not start from this
function, but from a symplectic $(2n)$- or $(2n+2)$-dimensional complex
space.
Symplectic transformations lead either to isometries on the manifold
or to symplectic reparametrizations. Finally we touch on the
connection with special \qu\ and very special real manifolds, and the
classification of homogeneous special manifolds.
\vfill      \hrule width 5.cm
\vskip 2.mm
{\small\small
\noindent $^\dagger$ Based on invited lectures delivered by AVP
at the Spring workshop on String theory, Trieste,
April 1994; to be published in the proceedings.\\
$^*$ Onderzoeksleider, N.F.W.O., Belgium}
\end{quote}
\begin{flushleft}
October 1995
\end{flushleft}
\end{titlepage}
\begin{abstract}
\end{abstract}
\maketitle
\setcounter{footnote}{0}
\section{Introduction}
In nonlinear sigma models, the spinless fields define a map from the
$d$-dimensional Minkowskian space-time to some `target space',
whose metric is given by the kinetic terms of these scalars.
Supersymmetry severely restricts the
possible target-space geometries. The type of target space which one
can obtain depends on $d$ and on $N$, the latter indicating the number of
independent supersymmetry transformations. The number of
supersymmetry generators (`supercharges') is thus equal to $N$ times
the dimension of the (smallest) spinor representation. For realistic
supergravity this number of supercharge components cannot exceed 32.
As 32 is the number of components of a Lorentz spinor in $d=11$
space-time dimensions, it follows that realistic supergravity
theories can only exist for dimensions $d\leq 11$. For the physical
$d=4$ dimensional space-time, one can have supergravity theories
with $1\leq N\leq 8$.
\par
\begin{table*}[hbt]
\setlength{\tabcolsep}{1.5pc}
\newlength{\digitwidth} \settowidth{\digitwidth}{\rm 0}
\catcode`?=\active \def?{\kern\digitwidth}
\caption{Restrictions on target-space manifolds
according to the type of supergravity theory. The rows are arranged
such that the number $\kappa$ of supercharge components is constant.
${\cal M}$ refers to a general Riemannian
manifold, $SK$ to `special \Ka', $VSR$ to
`very special real' and $Q$ to \qu\ manifolds. }
\label{tbl:mandN}
\begin{tabular}{c|ccccc}\hline
$\kappa$& $d=2$ & $d=3$ & $d=4$ & $d=5$ & $d=6$  \\ \hline
&$N=1$     &&&& \\
2&${\cal M}$ &&&&  \\   \hline
&$N=2$ &  $N=2$  & $N=1$ &   &  \\
4&\Ka   &  \Ka  & \Ka   &    &    \\ \hline
&$N=4$ &  $N=4$ & $N=2$        & $N=2$    & $N=1$     \\
8&$Q $  &   $Q$   &  $SK\oplus Q$ & $VSR\oplus Q$ & $\O \oplus Q $   \\
\hline  \hline
&...    & ... & $N=4$  & ... & $\rightarrow $     \\
16&...    & ... & $\frac{SO(6,n)}{SO(6)\otimes SO(n)}\otimes
\frac{SU(1,1)}{U(1)}$  & ... & $d=10$\\ \hline
&...    & ... & $N=8$  & ... &   $\rightarrow $ \\
32&...    & ... & $\frac{E_7}{SU(8)}$  & ... & $d=11$ \\ \hline
\end{tabular}
\end{table*}
\par
As clearly exhibited in table~\ref{tbl:mandN}, the more supercharge
components one has, the more restrictions one finds. When the number
of supercharge components exceeds 8, the target spaces are
restricted to symmetric spaces. For $\kappa=16$ components, they
are specified by an integer $n$, which specifies the number of
vector multiplets. This row continues to $N=1$, $d=10$. Beyond 16
supercharge
components there is no freedom left. The row with 32 supercharge
components continues to $N=1$, $d=11$. Here we treat the case
of 8 supercharge components. This is the highest value of $N$
where the target space is not yet restricted to be a symmetric
space, although supersymmetry has already
fixed a lot of its structure. We will
mostly be concerned with $N=2$ in $d=4$ dimensions. The
target space factorizes into a \qu\ and a \Ka\ manifold of a
particular type \cite{DWVP}, called {\em special}
\cite{special}. The former contains the scalars of the hypermultiplets
(multiplets without vectors). The latter contain the scalars in
vector multiplets. Recently the special
\Ka\ structure received a lot of attention, because it plays an
important role in string compactifications. Also \qu\
manifolds appear in this context, and also here it is a
restricted class of special
\qu\ manifolds that is relevant. In lowest order of the string
coupling constant these manifolds are even `very special' \Ka\
and \qu, a notion that we will define below.
\par
In the next section we describe the actions of $N=2$ vector
multiplets. First we consider rigid supersymmetry. We explain
the fields in the multiplets, their description in superspace and
how this leads to a holomorphic prepotential. Then we exhibit
how the structure becomes more complicated in supergravity, where
the space of physical scalars is embedded in a projective space.
This became apparent by starting from the superconformal tensor calculus.
\par
In section~\ref{ss:sympltr} we discuss the symplectic
transformations, which play an important role in the recent
developments of weak--strong coupling dualities. First we repeat
the general idea (and elucidate it for $S$ and $T$ dualities),
and then show what is the extra structure in $N=2$ theories. There
are two kind of applications, either as isometries of the manifolds
(symmetries of the theory), or as equivalence relations of
prepotentials (pseudo-symmetries). We illustrate both with
explicit examples. These will also exhibit formulations without a
prepotential, showing the need for a formulation that does not
rely on the existence of a
prepotential. This formulation is given at the end of the section.
Some further results will be mentioned in section~\ref{ss:further}.
\par
In all of this we confine ourselves to special
geometry from a supersymmetry/supergravity perspective. The
connection with the geometry of the moduli of Calabi-Yau
spaces \cite{special,Seiberg,CecFerGir,FerStro,DixKapLou,Cand}
is treated in the lectures of Pietro Fr\`e \cite{lectPietro}.
\section{$N=2$ actions} \label{ss:N2act}
\begin{table}[hbt]  \label{tbl:multN2d4}
\catcode`?=\active \def?{\kern\digitwidth}
\caption{Physical fields in $N=2$, $d=4$ actions}
\begin{tabular}{|c|ccc|}\hline
spin& pure SG &$n$ vector m. &$s$ hyperm. \\ \hline
2& 1& & \\
$3/2$& 2& & \\
1&1 &n & \\
$1/2$&&2n & 2s \\
0&&2n& 4s \\ \hline
\end{tabular}
\end{table}
We briefly introduce special \Ka\ manifolds in the context of
$N=2$, $d=4$ supergravity. As exhibited in table~\ref{tbl:multN2d4},
the physical multiplets of supersymmetry are vector and
hypermultiplets, which can be coupled to supergravity. In this
section we will not consider the hypermultiplets.
The scalar sector of the $N=2$ supergravity-Yang-Mills
theory in four space-time dimensions defines the `special \Ka\
manifolds'. Without supergravity we have $N=2$ supersymmetric
Yang-Mills theory, which we will treat first. The spinless fields
parametrize then a similar type of \Ka\ manifolds. The vector
potentials, which describe the spin-1 particles, are accompanied by
complex scalar fields and doublets of spinor fields, all taking
values in the Lie algebra associated with the group that can be
gauged by the vectors. In the second subsection we will see what
the consequences are of mixing the vectors in the vector
multiplets with the one in the supergravity multiplet.
\subsection{Rigid supersymmetry}
The superspace contains the anticommuting coordinates
$\theta^i_\alpha$ and $\bar \theta_{\dot \alpha i}$ where
    $i=1,2$ and $\alpha, \dot \alpha$ are the spinor indices.
The simplest superfields are, as in $N=1$, the chiral superfields.
They are defined by a constraint $ \bar D^{\dot \alpha i}\Phi=0 $,
where $\bar D^{\dot \alpha }$ is a covariant chiral superspace
derivative, and $\Phi$ is a complex superfield. This constraint
determines its structure\footnote{We use
${\cal F}^\pm_{\mu\nu}= \ft12\left({\cal F}_{\mu\nu}\pm
\ft12 \epsilon _{\mu\nu\rho\sigma}{\cal F}^{\rho\sigma}\right) $ with
$\epsilon^{0123}=i$.}:
\begin{eqnarray}
\Phi&=& X + \theta^i_\alpha  \lambda_i^\alpha  \label{chircomp} \\ &&+
\epsilon_{ij} \theta^i_\alpha \sigma^{\alpha\beta} _{\mu\nu}
\theta^j_\beta {\cal F}^{+\mu\nu} +
\epsilon_{\alpha\beta} \theta^i_\alpha  \theta^j_\beta Y_{ij}
+\ldots \ ,   \nonumber
\end{eqnarray}
where $\ldots $ stands for terms cubic or higher in $\theta$. New
component fields can appear up to $\theta^4$, leading to
$8+8$ complex field components. All these fields do not form an irreducible
representation of supersymmetry, but can be split into two sets of
$8+8$ real fields transforming irreducibly. We restrict ourselves
to the
set containing the fields already exhibited in \eqn{chircomp}, which
leads to the vector multiplet (The others form a `linear
multiplet'). The reduction is accomplished by the additional constraint
\begin{equation}
D_{(i}^\alpha D_{j)}^\beta \Phi\, \epsilon_{\alpha\beta} =
\epsilon_{ik}\epsilon_{j\ell } \bar D^{\dot \alpha(k}\bar D^{\ell )\dot
\beta}\bar \Phi \, \epsilon_{\dot \alpha\dot \beta} \ ,
\end{equation}
which for instance implies  that the symmetric tensor $Y_{ij}$
satisfies a  reality constraint:
$Y_{ij}=\epsilon_{ik}\epsilon_{j\ell }\bar Y^{k\ell }$, so that
it consists of only 3 real scalar fields. But more
importantly, we also obtain a constraint on the antisymmetric tensor:
$\partial^\mu\left( {\cal F}^+_{\mu\nu}-{\cal
F}^-_{\mu\nu}\right)=0$,
which is the Bianchi identity, which implies that ${\cal F}$ is
the field strength of a
vector potential. All the terms $\ldots $ in \eqn{chircomp} are
determined in terms of the fields written down. Therefore the
independent components of the vector multiplet are: $X^A,
\lambda^{iA}, {\cal F}^A_{\mu\nu}, Y_{ij}^A $ (where $A=1,..., n$
denotes the possibility to include several multiplets). $X^A$ and
$\lambda^{iA}$ will describe the physical scalars and spinors, ${\cal
F}^A$ are the fields strengths of the vectors and $Y^A$ will be
auxiliary scalars in the actions which we will construct.

As we have a chiral superfield, an action can be obtained by
integrating an arbitrary holomorphic function $F(\Phi)$ over chiral
superspace. The action
\begin{equation}
\int d^4x\int d^4\theta\ F(\Phi)\ + c.c.   \label{superspaceF}
\end{equation}
leads to the Lagrangian
\begin{eqnarray}
{\cal L} &=& g_{A \bar B}\partial_\mu X^A \partial_\mu \bar X^B+
 g_{A\bar B} \bar\lambda^{i A} \dsl \lambda^{\bar B }_i+
\\ && + \Im (F_{AB}{\cal F}_{\mu\nu}^{-A}{\cal F}_{\mu\nu}^{-B})
+{\cal L}_{\rm Pauli}+{\cal L}_{\rm 4-fermi}  \nonumber
\end{eqnarray}
where the latter two terms are the couplings of the vector fields to
the spinors and the terms quartic in fermions, which we do not write
explicitly here. The metric in target space is K\"ahlerian:
\cite{PKTN2}
\begin{eqnarray}
g_{A \bar B}(X,\bar X) & = &\partial_A\partial_{\bar B}K(X,\bar
X)\label{Karigid}\\
K(X,\bar X)&=&i (\bar F_A(\bar X) X^A-F_A(X) \bar X^A) \nonumber\\
F_A(X) &=&\partial_A
F(X)\ ;\ \bar F_A(\bar X)=\partial_{\bar A} \bar F(\bar X)\ . \nonumber
\end{eqnarray}
For $N=1$ the \Ka\ potential could have been arbitrary. The
presence of two independent supersymmetries
implies that this \Ka\ metric, and even the complete action,
depends on
a holomorphic prepotential $F(X)$, where $X$ denotes the complex
scalar fields.
Two different functions $F(X)$ may correspond to equivalent
equations of motion and to the same geometry.
{}From the equation\footnote{%
    Here and henceforth we use the convention where
    $F_{AB\cdots}$ denote multiple derivatives with respect to $X$ of
    the holomorphic prepotential. } $g_{A \bar B} =2\,\Im F_{AB}$,
it follows that
\begin{equation}
F\approx F +a+ q_A X^A + c_{AB}X^A X^B \ ,\label{FapprF}
\end{equation}
where $a$ and $q_A$ are complex numbers, and $c_{AB}$ real\footnote{
In supergravity, or in the full quantum theory the $q_A$ must be
zero.}.
But more relations can be derived from the symplectic
transformations that we discuss shortly.

The fact that the metric is \Ka ian implies that only curvature
components with two holomorphic and two anti--holomorphic indices can
be non--zero. In this case, these are determined by the third
derivative of $F$:
\begin{equation}
R_{A \bar B C \bar D}=- F_{ACE}  g^{E \bar F} \bar F_{FBD}\ .
\label{Rrigid}
\end{equation}
\subsection{Vector multiplets coupled to supergravity }
\label{ss:vectormN2SG}
The general action for vector multiplets coupled to $N=2$
supergravity was first derived using superconformal tensor calculus
\cite{DWVP}. In that approach one starts from the $N=2$
superconformal group, which is
\begin{equation}
SU(2,2|N=2)\supset SU(2,2)\otimes U(1)\otimes SU(2)\ .
\end{equation}
The bosonic subgroup, which we exhibited, contains, apart from the
conformal group in $d=4$, also $U(1)$ and $SU(2)$ factors. The \Ka
ian nature of vector multiplet couplings and the \qu\ nature of
hypermultiplet couplings is directly related to the presence of
these two groups. The
superconformal group is, however, mainly a useful tool for
constructing
actions which have just super-Poincar\'e invariance (see the reviews
\cite{revN2}). To make that transition, the dilatations, special
conformal transformations and $U(1)\otimes SU(2)$ are broken by an
explicit gauge fixing. The same applies to some extra
$S$--supersymmetry in the fermionic sector.

To describe theories as exhibited in table~\ref{tbl:multN2d4}, the
following multiplets are introduced: (other possibilities, leading
to equivalent physical theories, also exist, see
\cite{dWLVP,revN2}). The {\em Weyl multiplet} contains the vierbein,
the two gravitinos, and auxiliary fields. We introduce {\em $n+1$
vector multiplets}~:
\begin{equation}
\left( X^I,\lambda^{iI},
{\cal A}_\mu^I  \right)\qquad\mbox{with}\qquad I=0,1,...,n.
\end{equation}
The extra vector multiplet labelled by $I=0$ contains the scalar
fields which are to be gauge--fixed in order to break dilations and
the $U(1)$, the fermion to break the $S$--supersymmetry, and the
vector which corresponds to the physical vector of the supergravity
multiplet in table~\ref{tbl:multN2d4}.
Finally, there are {\it $s+1$ hypermultiplets}, one of these
contains only auxiliary fields and fields used for the gauge fixing
of $SU(2)$.
For most of this paper we will not discuss hypermultiplets ($s=0$).

Under dilatations the scalars $X^I$ transform with weight~1. On the
other hand an action similar to \eqn{superspaceF} can only be
constructed if $F(X)$ has Weyl weight~2. This leads to the important
conclusion that for the coupling of vector multiplets to
supergravity, one again starts from a holomorphic prepotential $F(X)$,
this time of $n+1$ complex fields, but now it
must be a {\em homogeneous} function of degree two  \cite{DWVP}.

In the resulting action appears
$-\ft12 i(\bar X^I F_I -X^I\bar F_I)eR$, where $R$ is the space--time
curvature. To have the canonical kinetic terms for the graviton, it
is therefore convenient to impose as gauge fixing for dilatations the
condition
\begin{equation}
i(\bar X^I F_I - \bar F_I X^I) = 1\,.\label{constraint}
\end{equation}
Therefore, the physical scalar fields
parametrize an $n$-dimensional complex hypersurface,
defined by the condition \eqn{constraint}, while the overall phase of the
$X^I$ is irrelevant in view of a local (chiral) invariance.
 The embedding of this hypersurface can be described in terms of
$n$ complex coordinates $z^A$ by letting $X^I$ be proportional to
some holomorphic sections $Z^I(z)$ of the projective space
$P\Cbar^{n+1}$ \cite{CdAF}.
The bosonic part of the resulting action is (without gauging)
\begin{eqnarray}
e^{-1}{\cal L}&=&-\ft12 R+ g_{\alpha \bar \beta }\partial_\mu z^\alpha
\partial^\mu \bar z^{\bar \beta }  \nonumber\\ &&
- \Im\left( {\cal N}_{IJ} (z,\bar z) {\cal F}_{\mu\nu}^{+I}
{\cal F}_{\mu\nu}^{+J} \right).
\end{eqnarray}

The $n$-dimensional space parametrized by the
$z^\alpha $ ($\alpha =1,\ldots, n$) is a \Ka\ space; the K\"ahler metric
$g_{\alpha \bar \beta }=\partial_\alpha \partial_{\bar \beta }
K(z,\bar z)$ follows from the K\"ahler potential
\begin{eqnarray}
&&e^{-K(z,\bar z)}=
i \bar Z^I(\bar z)\,F_I(Z(z)) -i Z^I(z)\,
\bar F_I(\bar Z(\bar z))\nonumber\\
&& X^I=e^{K/2}Z^I(z)\ ,\qquad\bar X^I=e^{K/2}\bar Z^I(\bar z) \ .
 \label{Kalocal}
\end{eqnarray}
The resulting geometry is known as {\em special} \Ka\ geometry
\cite{DWVP,special}. The  curvature tensor associated with this
K\"ahler space satisfies the characteristic relation
\cite{BEC}
\begin{equation}
R^\alpha {}_{\!\!\beta \gamma }{}^{\!\delta } = \d^\alpha _{\beta }
\d^\delta _{\gamma }+\d^\alpha _{\gamma } \d^\delta _{\beta } -
e^{2K} {\cal W}_{\beta \gamma \epsilon }\, \bar {\cal W}{}^{\epsilon
\alpha \delta }\, , \label{SKcurvature}
\end{equation}
where
\begin{equation}
{\cal W}_{\alpha \beta \gamma } = i F_{IJK}\big(Z(z)\big) \;{\partial
  Z^I\over \partial z^\alpha }  {\partial Z^J\over \partial z^\beta }
{\partial Z^K\over \partial z^\gamma } \,.
\end{equation}
\par
A convenient choice of inhomogeneous coordinates $z^\alpha $
are the {\em special} coordinates, defined by
\begin{equation}
z^A =X^A/X^0,\qquad A=1,\ldots ,n, \label{defspcoor}
\end{equation}
or, equivalently,
\begin{equation}
Z^0(z)=1\,,\qquad Z^A(z) = z^A\,.
\end{equation}
\par
The kinetic terms of the spin-1 gauge fields in the action are
proportional to the symmetric tensor
\begin{equation}
{\cal N}_{IJ}=\bar
F_{IJ}+2i {{\rm Im}(F_{IK})\,{\rm Im}(F_{JL})\,X^KX^L\over {\rm
Im}(F_{KL})\,X^KX^L} \,.
\label{Ndef}
\end{equation}
This tensor describes the field-dependent generalization of the inverse
coupling constants and so-called $\theta$ parameters.

We give here some examples of functions $F(X)$ and their
corresponding target spaces, which will be useful later on:
\begin{eqnarray}
F=-i\,X^0X^1  &\quad& \frac{SU(1,1)}{U(1)}  \label{FiX0X1}\\
 F=(X^1)^3/X^0  &\qquad&  \frac{SU(1,1)}{U(1)}  \label{FX13}\\
 F=-4\sqrt{X^0(X^1)^3}   &&  \frac{SU(1,1)}{U(1)}  \label{Fsqrt} \\
F=iX^I\eta_{IJ}X^J &&
\frac{SU(1,n)}{SU(n)\otimes U(1)}\\
 F=\frac{d_{ABC} X^A X^B X^C}{X^0} &&  \mbox{`very special'} \label{Fvsp}
\end{eqnarray}
The first three functions give rise to the manifold $SU(1,
1)/U(1)$. However,
the first one is not equivalent to the other two as the
manifolds have a different value of the curvature \cite{CremVP}.
The latter two are, however, equivalent by means of a symplectic
transformation as we will show below. In the fourth example
$\eta$ is a constant non-degenerate real symmetric matrix. In
order that the manifold has a
non-empty positivity domain, the signature of this matrix should be
$(+-\cdots -)$. So not all functions $F(X)$ allow a non-empty
positivity domain. The last example, defined by a real symmetric tensor
$d_{ABC}$, defines a class of special \Ka\ manifolds, which we will
denote as `very special' \Ka\ manifolds. This class of manifolds
is important in the applications discussed below.

\section{Symplectic transformations} \label{ss:sympltr}
The symplectic transformations are a generalization of the
electro-magnetic duality transformations. We first recall the
general formalism for arbitrary actions with coupled spin-0 and
spin-1 fields, and then come to the specific case of $N=2$.
\subsection{Pseudo-symmetries in general}
We consider general actions of spin-1 fields with field strengths
${\cal F}_{\mu\nu}^\Lambda$ (now labelled by $\Lambda=1, ..., m$) coupled to
scalars. The general form of the kinetic terms of the spin 1 fields
is
\begin{eqnarray}
{\cal L}_1&=&
\ft14 (\Im {\cal N}_{\Lambda\Sigma}){\cal F}_{\mu\nu}^\Lambda
{\cal F}^{\mu\nu\Sigma}
\nonumber\\ &&
-\ft i8 (\Re {\cal N}_{\Lambda\Sigma})
\epsilon^{\mu\nu\rho\sigma}{\cal F}_{\mu\nu}^\Lambda
{\cal F}_{\rho\sigma}^\Sigma
\nonumber\\ &=&
\ft12 \Im \left({\cal N}_{\Lambda\Sigma}
{\cal F}_{\mu\nu}^{+\Lambda} {\cal F}^{+\mu\nu\Sigma}\right)
\end{eqnarray}
We define
\begin{eqnarray}
  G_{+\Lambda }^{\mu\nu}\equiv 2i\frac{\partial{\cal L}}
  {\partial \cF^{+\Lambda }_{\mu\nu}}=
{\cal N}_{\Lambda\Sigma}\cF^{+\Sigma \,\mu\nu}
\nonumber\\
  G_{-\Lambda }^{\mu\nu}\equiv -2i\frac{\partial{\cal L}}
  {\partial \cF^{-\Lambda }_{\mu\nu}}=
\bar{\cal N}_{\Lambda \Sigma }\cF^{-\Sigma \,\mu\nu}\ .\label{defG}
\end{eqnarray}
The equations for the field strengths can then be written as
\begin{eqnarray*}
\del^\mu \Im \cF^{+\Lambda }_{\mu\nu} &=&0\ \ \ \ \ {\rm Bianchi\
identities}\\
\del_\mu \Im G_{+\Lambda }^{\mu\nu} &=&0\ \ \ \ \  {\rm Equations\  of\
motion}
\end{eqnarray*}
This set of equations is invariant under  $GL(2m,\Rbar)$
transformations:
\begin{equation}
\pmatrix{\widetilde\cF^+\cr \widetilde G_+\cr}={\cal S}
\pmatrix{\cF^+\cr G_+\cr} =
\pmatrix{A&B\cr C&D\cr}   \pmatrix{\cF^+\cr G_+\cr}\ . \label{FGsympl}
\end{equation}
However, the $G_{\mu\nu}$ are related to the ${\cal F}_{\mu\nu}$
as in \eqn{defG}. The previous transformation implies
\begin{eqnarray}
\widetilde G^+&=&(C+D{\cal N})F^+\nonumber\\ &=&
(C + D{\cal N})(A+B{\cal N})^{-1} \widetilde F^+ \ .
\end{eqnarray}
Therefore the new tensor ${\cal N}$ is
\begin{equation}
\mbox{ \parbox[t]{3.8cm}
{\fbox{$\widetilde{\cal N} = (C + D{\cal N})(A+B{\cal N})^{-1}$}}}
\label{tilNN}\end{equation}
This tensor should be symmetric, as it is the second derivative of
the action with respect to the field strength. This request
leads to the equations which determine that
${\cal S}\in Sp(2m,\Rbar)$, i.e.
\begin{eqnarray}
&& {\cal S}^T  \Omega   {\cal S}   =  \Omega  \qquad\mbox{where}\qquad
\Omega=\pmatrix{0&\bigone \cr -\bigone &0\cr} \nonumber\\
&&\mbox{or }\left\{\begin{array}{l}
A^T C-C^T A=0\\ B^T D- D^T B=0\\ A^T D-C^T B=\bigone
\end{array}\right.
\label{Ssympl}\ .
\end{eqnarray}
Some remarks are in order: First, these transformations act on
the field strengths. They generically rotate electric into magnetic fields
and vice versa. Such rotations, which are called duality
transformations, because in four space-time dimensions electric
and magnetic fields are dual to each other in the sense of
Poincar\'e duality, cannot be implemented on the vector potentials,
at least not in a local way. Therefore, the
use of these symplectic transformations is only legitimate for zero
gauge coupling constant. From now on, we deal
exclusively with Abelian gauge groups. Secondly, the Lagrangian is
not an invariant if $C$ and $B$ are not zero:
\begin{eqnarray}
&&\Im \widetilde\cF^{+\Lambda}  \widetilde G_{+\Lambda}   =
  \Im \left(\cF^{+}  G_{+ }\right)  \nonumber\\
  &&+\Im \left(2 \cF^{+} (C^T B) G_+
 + \cF^{+ }(C^T A) \cF^{+} \right. \nonumber\\ &&\left.
 +G_{+ } (D^T B) G_{+ } \right)\ .
\end{eqnarray}
If $C\neq 0, B=0$ it is invariant up to a four--divergence. Thirdly,
the transformations can also act on dyonic solutions of the field
equations and the vector $\pmatrix{q^\Lambda_m\cr q_{e\,\Lambda}\cr}$
of magnetic and electric charges transforms also as a symplectic
vector. The Schwinger-Zwanziger quantization condition restricts
these charges to a lattice with minimal surface area proportional
to $\hbar$. Invariance of this lattice restricts the symplectic
transformations to a discrete subgroup:
\begin{equation}
{\cal S}\in Sp(2m,\IZ) \ .
\end{equation}
Finally, the transformations with $B\neq 0$ will be
non--perturbative. This can be seen from the
fact that they do not leave the purely electric charges
invariant, or from the fact that \eqn{tilNN} shows that these
transformations invert ${\cal N}$ which plays the role of the gauge
coupling constant.
\subsection{Pseudo--symmetries and proper symmetries}
The transformations described above, change the matrix ${\cal N}$,
which are gauge coupling constants of the spin-1 fields. This
can be compared to diffeomorphisms of the scalar manifold
$z\to \hat z(z)$ which change the metric (which is the coupling
constant matrix for the kinetic energies of the scalars) and
${\cal N}$:
\[
\hat g_{\alpha \beta } (\hat z(z)) {\partial\hat z^\alpha
\over \partial  z ^\gamma }
{\partial\hat z ^\beta  \over \partial  z ^\delta }=
g_{\gamma \delta } (z)\ ;
\hat {\cal N}(\hat z(z))={\cal N}(z)\ .
\]
Both these diffeomorphisms and symplectic reparametrizations
are {\em `Pseudo--symmetries':} \cite{christoi}
\begin{equation}
  D_{pseudo}=Diff({\cal M})\times Sp(2m,\Rbar)\ .
\end{equation}
They leave the action form invariant, but change the coupling
constants and are thus not invariances of the action.

If $\hat g_{\alpha \beta }(z)=g_{\alpha \beta }(z)$ then the
diffeomorphisms become
isometries of the manifold, and proper symmetries of the scalar
action. If these isometries are combined
with symplectic transformations such that
\begin{equation}
\hat {\widetilde{\cal N}}(z)={\cal N}(z)\ ,
\end{equation}
then this is a {\it proper symmetry}. These are
invariances of the equations of motion (but not
necessarily of the action as not all transformations can be
implemented locally  on the gauge fields).
To extend the full group of isometries of the
scalar manifold to proper symmetries, one thus has to embed this
isometry group in $Sp(2m;\Rbar)$, and arrives at the following
situation:
\[
D_{prop}=Iso(\cM)  \subset Iso(\cM)\times Iso(\cM) \subset D_{pseudo}
\]

Let us illustrate how $S$ and $T$ dualities, treated in Sen's
lectures \cite{Senhere}, fit in this scheme as proper
symmetries. The action he treats occurs in $N=4$ supergravity. The
scalars are $\lambda=\lambda_1 +i \lambda_2$ and a symmetric matrix
$M$, satisfying $M\eta M=\eta ^{-1}$ where $\eta =\eta ^T$ is the
metric of $O(6,22)$. Their coupling to the spin-1 fields is encoded
in the matrix
\begin{equation}
{\cal N}= \lambda_1 \eta+i\lambda_2\eta M\eta\ .
\end{equation}
The transformations on the scalars should lead to \eqn{tilNN} with
\eqn{Ssympl}. Let us first consider this for the {\em $T$ dualities}.
These are transformations of $O(6,22)$:
\begin{equation}
\widetilde{\cal F}^+ = A{\cal F}^+ \ ;\qquad
\widetilde M = A M A^T\ ,
\end{equation}
($\lambda$ is invariant) where $\eta =A^T \eta  A$. This leads to
$\widetilde{\cal N}= \left( A^T\right) ^{-1}{\cal N}A^{-1}$, which is
of the form \eqn{tilNN}, identifying $D=\left( A^T\right) ^{-1}$. The
matrices $C$ and $B$ are zero, which indicates that these symmetries are
realised perturbatively.

For the {\em $S$ dualities}, $M$ is invariant. These transformations
are determined by the integers $s,r,q,p$ such that $sp-qr=1$:
\[
\widetilde{\cal F}^+=  s{\cal F}^++ r \eta ^{-1}{\cal N}{\cal F}^+\ ;\
\widetilde \lambda= \frac{p\lambda + q}{r\lambda +s}\ .
\]
This leads to $\widetilde{\cal N}=
(p{\cal N}+q\eta )(r\eta ^{-1}{\cal N}+s)$, which is of the required
form upon the identification
\begin{equation}
{\cal S}= \pmatrix{s\bigone & r\eta ^{-1}\cr q\eta &p\bigone }\ .
\label{cSSU11}\end{equation}
Now, $B$ and $C$ are non-zero, which shows the non-perturbative
aspect of the $S$-duality.

\subsection{Symplectic transformations in $N=2$}
In $N=2$ the tensor ${\cal N}$ is determined by the function $F$ as
explained in section~\ref{ss:N2act}. The definitions of ${\cal N}$ in
rigid and local supersymmetry can be written in a clarifying way as
follows\footnote{%
For the rigid case, here $\partial_{\bar C}\bar X^B =\delta^B_C$,
but this definition is also applicable when we take derivatives
w.r.t. arbitrary coordinates $z^\alpha (X)$. For the local case one
regards $(\partial_{\bar \gamma } \bar F_I, F_I)$ as an $n+1$ by $n+1$
matrix to see how this defines the matrix ${\cal N}$.}
\begin{eqnarray}
\mbox{rigid SUSY}&\qquad&\mbox{SUGRA}\nonumber\\
\partial_{\bar C} \bar F_A={\cal N}_{AB}\partial_{\bar C}\bar X^B& &
\partial_{\bar \gamma } \bar F_I={\cal N}_{IJ}
\partial_{\bar \gamma }\bar X^J\\
&&   F_I= {\cal N}_{IJ}   X^J  \nonumber
\end{eqnarray}
{}From this definition it is easy to see that ${\cal N}$ transforms
in the appropriate way if we define
\begin{eqnarray}
 V=\pmatrix {X^A\cr F_A} &\qquad&  V=\pmatrix {X^I\cr F_I}
 \nonumber\\
 U_C =\pmatrix {\partial_C X^A\cr \partial_C F_A} &\qquad &
U_\alpha =\pmatrix {\partial_\alpha X^I\cr \partial_\alpha F_I}
\label{defVU}
\end{eqnarray}
(and their complex conjugates) as symplectic vectors in the two
cases. They thus transform as in \eqn{FGsympl}. With this
identification in mind, we can reconsider the kinetic terms of the
scalars. Then it is clear that the \Ka\ potentials \eqn{Karigid} and
\eqn{Kalocal}, and the constraint \eqn{constraint} are symplectic
invariants. This will lead to a new formulation of special geometry
in section~\ref{ss:sympldef}.

When we start from a prepotential $F(X)$, the $F_I$ are the
derivatives\footnote{The remarks below are written with indices $I,J$
as in the supergravity case, but can be applied as well in rigid
supersymmetry replacing these indices by $A,B$.} of $F$.
The expression $\tilde X^I= A^I{}_J X^J + B^{IJ} F_J(X)$ expresses
the dependence of the new coordinates $\tilde X$ on the old
coordinates $X$. If this transformation is invertible \footnote{The full
symplectic matrix is always invertible, but this part may not
be. In rigid supersymmetry, the invertibility of this transformation
is necessary for the invertibility of ${\cal N}$, but in supergravity
we may have that the $\tilde X^I$ do not form an independent set, and
then $\tilde F$ can not be defined. See below.},  the $\tilde F_I$
are again the derivatives of an new function $\tilde F(\tilde X)$
of the new coordinates,
\begin{equation}
\tilde F_I(\tilde X)=\frac{\partial \tilde F(\tilde X)}
{\partial \tilde X^I}\ .
\end{equation}
The integrability condition which implies this statement is equivalent
to the condition that ${\cal S}$ is a symplectic matrix. In the
supergravity case, one can obtain $\tilde F$ due to the homogeneity:
\begin{equation}
\tilde F(\tilde X(X))=
\frac{1}{2} V^T \pmatrix{C^T A& C^T B \cr D^T A & D^T B\cr}V\ .
\label{tilFexpl}
\end{equation}
Hence we obtain a new formulation of the theory, and thus of the
target-space manifold, in terms of the function $\tilde F$.

We have to distinguish two  situations:\\
1. The function $\tilde F(\tilde X)$ is different from $F(\tilde X)$,
even taking into account \eqn{FapprF}.
In that case the two functions describe
equivalent classical field theories. We have a {\em pseudo symmetry}.
These transformations are
called symplectic reparametrizations \cite{CecFerGir}.
Hence we may find a variety of descriptions of the same theory
in terms of different functions  $F$. \\
2. If a symplectic transformation leads to the same function $F$
(again up to \eqn{FapprF}), then we are dealing with a {\em proper
symmetry}. As explained above, this invariance reflects itself in
an isometry of the target-space manifold.
Henceforth these symmetries are called
`duality symmetries', as they are generically accompanied by
duality transformations on the field equations and the Bianchi
identities. The question remains whether the
duality symmetries comprise all the isometries of the target
space, i.e. whether
\begin{equation}
Iso({\cal M})\subset Sp(2(n+1),\Rbar)\ . \label{isosubSp}
\end{equation}
We investigated this question in \cite{brokensi} for the very
special \Ka\ manifolds, and found that in that case one does obtain the
complete set of isometries from the symplectic transformations.
For generic special \Ka\ manifolds no isometries have been found
that are not induced by symplectic
transformations, but on the other hand there is no proof that
these do not exist.

\subsection{Examples (in supergravity)}
We present here some examples of symplectic reparametrizations and
duality symmetries in the context of $N=2$ supergravity. First
consider \eqn{FX13}. If we apply the symplectic transformation
\begin{equation}
{\cal S}=\pmatrix{A&B\cr C&D}=\pmatrix{1&0&0&0\cr 0&0&0&1/3\cr 0&0&1&0\cr
0&-3&0&0}
\end{equation}
one arrives, using \eqn{tilFexpl}, at \eqn{Fsqrt}. So this is
a symplectic reparametrization, and shows the equivalence of the two
forms of $F$ as announced above.

On the other hand consider
\begin{equation}
{\cal S}=\pmatrix{1+3\epsilon&\mu    &0&0\cr
           \lambda& 1+\epsilon &0&2\mu/9\cr
            0&0           &1-3\epsilon&-\lambda\cr
            0& -6\lambda &-\mu&1-\epsilon \cr }
\end{equation}
 for infinitesimal $\epsilon,\mu,\lambda$. Then $F$ is invariant. On
 the scalar field $z=X^1/X^0$, the transformations act as
\begin{equation}
\delta z=\lambda-2\epsilon z-\mu z^2/3 \ .
\end{equation}
They form an $SU(1,1)$ isometry group of the scalar manifold. The
domain were the metric is positive definite is $\Im z >0$. This shows
the identification of the manifold as the coset space in \eqn{FX13},
\eqn{Fsqrt}.

As a second example, consider \eqn{FiX0X1}. Using \eqn{Ndef} one
obtains the matrix ${\cal N}$ which determines (again with $z=X^1/X^0$)
\begin{equation}
e^{-1}{\cal L}_1= -\ft12 \Re\left[z \left( F_{\mu\nu}^{+0}\right) ^2
+z^{-1} \left( F_{\mu\nu}^{+1}\right) ^2 \right]\ .
\end{equation}
This appears also in pure $N=4$ supergravity in the so--called
`$SO(4)$ formulation' \cite{N4SO4}.  Consider now the symplectic mapping
\cite{f0art}
\begin{equation}
{\cal S}=\pmatrix{1&0&0&0\cr 0&0&0&-1\cr 0&0&1&0\cr 0&1&0&0\cr }\ ,
\end{equation}
leading to the transformations
\begin{eqnarray}
\widetilde X^0=X^0 &\qquad& \widetilde X^1=-F_1=iX^0\\
\widetilde F_0=F_0 &\qquad& \widetilde F_1= X^1\ .
\end{eqnarray}
This is an example where the transformation between $\tilde X$ and
$X$ is not invertible. Using \eqn{tilFexpl}, we obtain $\tilde
F=0$. However, $A+B{\cal N}$ is invertible, and we can compute
$\tilde N$ using \eqn{tilNN}, leading to
\begin{equation}
e^{-1}{\cal L}_1= -\ft12 \Re \left[z \left( F_{\mu\nu}^{+0}\right) ^2
+ z \left( F_{\mu\nu}^{+1}\right) ^2 \right]\ .
\end{equation}
(We performed here a symplectic transformation, but no
diffeomorphism. We are still using the same variable $z$). This is
the form familiar from the `$SU(4)$ formulation' of pure $N=4$
supergravity \cite{CSF}. This shows that there are formulations
which can not be obtained directly from a superspace action.

In the final example, we will show that this particular formulation
can be the most useful one. For that we consider the manifold
\begin{equation}
 \frac{SU(1,1)}{U(1)}\otimes \frac{SO(r,2)}{SO(r)\otimes SO(2)}\ .
\label{STr}\end{equation}
This is the only special \Ka\ manifold which is a product of two
factors \cite{splitspecial}. Therefore it appears in string theory
where the first factor contains the dilaton-axion. The first
formulation of this class of manifolds used a function $F$ of the
type \eqn{Fvsp}: $F(X) = \frac{1}{X^0} X^S X^r X^t \eta_{rt}$, where
 $\eta_{rt}$
is the constant diagonal metric with signature $(+, -, \ldots, -)$
\cite{BEC}.
In this parametrization only an $SO(r-1)$ subgroup of $SO(r,2)$ is
linearly realized (residing in $A$ and $D$ of \eqn{FGsympl}). From a string
compactification point of view one does not expect this. The full
$SO(r,2)$ should be a perturbative symmetry, as it is realized
in the $N=4$ theory described by Sen \cite{SenN4,Senhere}. In the
search for better parametrizations, by means of a symplectic
reparametrization a function $F$ of the square root type was
discussed in \cite{FreSoriani} which has $SO(r)$ linearly realized.
However, the solution was found in \cite{f0art}, and was not based on
a function $F$ at all. The symplectic vector $V$ contains then
\begin{equation}
F_I =S\,\eta_{IJ}\,X^J\ ,\label{FISXI}
\end{equation}
where $S$ is one of the coordinates
(representing the first factor of \eqn{STr}), and the $X^I$ satisfy
the constraint $ X^I\,\eta_{IJ}\,X^J =0$, where $\eta_{IJ}$ is the
$SO(2,r)$ metric. For additional details on this example, see also
\cite{dWKLL}, where the perturbative corrections to the vector multiplet
couplings are considered in the context of the $N=2$ heterotic
string vacua.
This important example shows that under certain
circumstances one needs a formulation that does not rely on the
existence of a function $F$.

\subsection{Coordinate independent description }
   \label{ss:sympldef}
We want to be able to use more general coordinates than the special
ones which appeared naturally in the superspace approach, and also to
set up a formulation of the theory in which the symplectic structure
is evident. First we will formulate this for {\bf the rigid case}
\cite{modssym}.

We start by introducing the symplectic vector $V\in \Cbar^{2n}$, as
in \eqn{defVU}, where now the $F_A$ are no longer the derivative
of a function $F$, but $n$ independent components. Then consider
functions $V(z)$, parametrized by $n$ coordinates $z^\alpha $
($\alpha =1, ...,
n$), which will be the coordinates on the special manifold. The
choice of special coordinates introduced before, corresponds to
$X^A(z)= z^\alpha , \, F_A(z)=\frac{\partial F}{\partial X^A}(X(z))$. By
taking now derivatives with respect to $z^\alpha $ one obtains $U_\alpha $
analogous to the $U_A$ in \eqn{defVU}.

We define as metric on the special manifold
\begin{equation}
g_{\alpha {\bar\beta}}=i\, U_\alpha ^T \, \Omega\, \bar
U_{\bar\beta} =i\, \langle U_\alpha , \bar U_{\bar\beta} \rangle \ ,
\end{equation}
where we introduced a symplectic inner product $\langle V,
W\rangle\equiv V^T \Omega W$. The constraints which define the rigid
special geometry can be formulated on the $2n\times 2n$ matrix
\begin{equation}
 {\cal V}\equiv \pmatrix{U_\alpha ^T \cr \bar U^{\alpha \,T}} \equiv
 \pmatrix{\del_\alpha  X^A &\del_\alpha  F_A\cr
g^{\alpha {\bar\beta}}\partial_{\bar\beta} \bar X^A &g^{\alpha
{\bar\beta}}\partial_{\bar\beta} \bar F_A}\ .
\end{equation}
This matrix should satisfy ${\cal V}\Omega{\cal V}^T=-i\Omega$ and
\[
{\cal D}_\alpha {\cal V}={\cal A}_\alpha {\cal V}
\qquad\mbox{with}\qquad
{\cal A}_\alpha =\pmatrix{0& C_{\alpha \beta \gamma } \cr 0&0\cr}
 \]
for a symmetric $C_{\alpha \beta \gamma }$ (being $F_{ABC}$ in
special coordinates);
and ${\cal D}$ contains the Levi-Civita connection. The integrability
condition of this constraint then implies the form of the curvature:
$R_{\alpha \bar \beta  \gamma  \bar \delta }=- C_{\alpha \gamma\epsilon}
\bar C_{\bar \beta  \bar \delta  \bar\epsilon} g^{\epsilon
\bar\epsilon}$ (compare this with \eqn{Rrigid}).
The formulation can even be simplified in terms of a
vielbein $e_\alpha ^A\equiv\del_\alpha  X^A $ (being the unit matrix in
special coordinates). Then the connection
$\hat \Gamma_{\alpha \beta }^\gamma =e^\gamma _A \partial_\beta
e_\alpha ^A$ is flat, and there are holomorphic constraints
\begin{eqnarray*}
&&\widehat{\cal V}\equiv\pmatrix{e_\alpha ^A &\del_\alpha F_A \cr 0
& e^\alpha _A\cr}
\nonumber\\
&&\partial_\alpha \widehat{\cal V} =\widehat {\cal A}_\alpha
\widehat{\cal V}
\qquad\mbox{with }
\qquad \widehat {\cal A}_\alpha =\pmatrix {\hat \Gamma_{\alpha \beta
}^\gamma & -iC_{\alpha \beta \gamma } \cr 0 & - \hat \Gamma_{\alpha
\gamma }^\beta }
\end{eqnarray*}

For {\bf Supergravity} a similar definition of special geometry is
possible. This formulation was first given in the context
of a treatment of the moduli space of Calabi-Yau three-folds
\cite{special,FerStro,Cand}. The particular way in which we present
it here is explained in more detail in \cite{prtrquat}. Now the
symplectic vectors have $2(n+1)$ components. We first impose the
constraint \eqn{constraint}, which is written in a symplectic way as
$ \langle \bar V,V\rangle \equiv  \bar V^T \Omega V =-i$. Then we
define $n$ holomorphic symplectic sections, parametrized by $z^\alpha $,
which are proportional to $V$:
\begin{equation}
V(z,\bar z)=  e^{\ft12 K(z, \bar z)}v(z) \ ,
\end{equation}
and the proportionality constant defines the \Ka\ potential.
These equations are then invariant under `\Ka\ transformations'
\begin{eqnarray}
&&v(z) \to e^{f(z)}\, v(z)\nonumber\\
&&K(z,\bar z)\to K(z,\bar z) - f(z) -\bar f(\bar z)\nonumber\\
&&V\to e^{\ft12(f(z)-\bar f(\bar z))}\,V\,. \label{Katransf}
\end{eqnarray}
for which $\partial_\alpha  K$ and $\partial_{\bar \alpha }K$ play the role
of connections.
Then special geometry is defined, using%
\footnote{The connection contains now the Levi--Civita one and the
\Ka\ connection related to \eqn{Katransf}: $\cD_\alpha X=\left( \del_\alpha
+{1\over2}(\del_\alpha K)\right) X$.} $U_\alpha ={\cal D}_\alpha V$,
with one additional constraint:
\begin{equation}
\langle U_\alpha ,U_\beta \rangle=0\ .
\end{equation}
Usually the $F_I(z)$ are functions which depend on $X^I(z)$. Then one
has $F_I=\partial_I F$, and the scaling symmetry implies that $F$ is
a holomorphic function homogeneous of 2nd degree in $X^I$. But e.g.
with \eqn{FISXI} this is not the case.

To make contact with the Picard-Fuchs equations in Calabi-Yau
manifolds, a similar formulation as for the rigid case is useful. This
is obtained by defining the $(2n+2)\times(2n+2)$ matrix
\begin{equation}
\cal V = \pmatrix{V\cr \bar U^\alpha  \cr \bar V\cr U_\alpha \cr}\ ,
\end{equation}
which satisfies ${\cal V}\,\Omega \,{\cal V}^{\rm T} = i\Omega$.
One then introduces a connection such that the constraints are
\cite{CDFLL}
\begin{eqnarray}
{\cal D}_\alpha {\cal V}&=& {\cal A}_\alpha {\cal V}\, , \qquad{\cal
D}_{\bar \alpha }
{\cal V}={\cal A}_{\bar \alpha }{\cal V}\,. \label{flatness}\\
\mbox{with e.g. }{\cal A}_\alpha  &=&
\pmatrix{0&0&0&\delta_\alpha ^\gamma \cr \noalign{\vskip1mm}
              0&0&\delta^\beta _\alpha &0\cr \noalign{\vskip1mm}
              0&0&0&0\cr \noalign{\vskip1mm}
              0& C_{\alpha \beta \gamma }&0&0\cr }\ .
\end{eqnarray}
The integrability conditions lead to the curvature tensor
\begin{equation}
R_{\alpha {\bar\beta} \delta \bar \gamma }=g_{\alpha {\bar\beta}}
g_{\delta \bar \gamma }+g_{\alpha \bar \gamma }g_{\delta
{\bar\beta}}-C_{\alpha \delta \epsilon}C_{{\bar\beta} \bar \gamma
\bar\epsilon} g^{\epsilon\bar \epsilon} \ .
\end{equation}
\section{Further results and conclusions}    \label{ss:further}
Special geometry is not confined to \Ka\ manifolds. There exist a
\cmap, which can be obtained either from dimensional reduction of
the field theory to 3 dimensions, or from superstring compactification
mechanisms \cite{CecFerGir}. This maps special \Ka\ manifolds
to a subclass of the \qu\
manifolds, which are then called special \qu. As already mentioned,
a subclass of special manifolds are the `very special' ones. These
can be obtained from dimensional reduction of actions in 5
dimensions, characterised by a symmetric tensor
$d_{ABC}$ \cite{GuSiTo}. This mapping is called the \rmap\
\cite{ssss}, and the manifolds in the 5-dimensional theory are
called `very special real' manifolds. These concepts were very
useful in the classification of homogeneous \cite{dWVP3} and
symmetric \cite{CremVP} special manifolds. It turned out that
homogeneous special manifolds are in one-to-one correspondence to
realizations of real Clifford algebras with signature $(q+1,1)$ for
real, $(q+2, 2)$ for \Ka, and $(q+3,3)$ for \qu\ manifolds. A
study of the full set of isometries could be done systematically
in these models . All this has been summarised in \cite{prtrquat}.

For string theory the implications of special geometry in the
rigid theories for the moduli spaces of Riemann surfaces
\cite{modssym}, and in the supergravity theories
for Calabi-Yau spaces
\cite{special,Seiberg,CecFerGir,FerStro,DixKapLou,Cand} is extremely
useful for obtaining non-perturbative results \cite{SeiWit,%
modssym,f0art}. For these results we refer to
\cite{lectPietro} and to \cite{fresoriabook}, where many more
aspects of special manifolds in the context of topological
theories, Landau-Ginzburg theories, etc. are discussed.\vspace{5mm}
\par
\noindent{ \bf Acknowledgements}\vspace{0.3cm}
\par
This review is a result of many collaborations. We want to thank all
our collaborators and especially  Sergio
Ferrara, Anna Ceresole, Riccardo D'Auria, Pietro Fr\`e, Marco Bill\'o
and Paolo Soriani.
\par
This work was carried out in the framework of the
project "Gauge theories, applied supersymmetry and quantum
gravity", contract SC1-CT92-0789 of the European Economic
Community.

\end{document}